\documentclass[10pt]{article}
\usepackage{bbm}
\usepackage[final]{graphics}
\usepackage{amsmath}
\usepackage{amsfonts,amsbsy}
\usepackage{amssymb}
\usepackage{graphicx}

\def\empile#1\over#2{\mathrel{\mathop{\kern 0pt#1}\limits_{#2}}}

\newcommand{\slcalP}{\raise.15ex\hbox{$/$}\kern-.63em\hbox{$\cal P$}}

\def\m{{\boldsymbol m}}

\catcode`\@=11

\newcount\@tempcntc
\def\@citex[#1]#2{\if@filesw\immediate\write\@auxout{\string\citation{#2}}\fi
  \@tempcnta\z@\@tempcntb\m@ne\def\@citea{}\@cite{%
        \@for\@citeb:=#2\do%
    {\@ifundefined{b@\@citeb}%
        {\@citeo\@tempcntb\m@ne\@citea%
                \def\@citea{,\penalty\@m\ }{\bf ?}\@warning%
                {Citation `\@citeb' on page \thepage \space undefined}}%
        {\setbox\z@\hbox{\global\@tempcntc0\csname b@\@citeb\endcsname\relax}
     \ifnum\@tempcntc=\z@ \@citeo\@tempcntb\m@ne%
       \@citea\def\@citea{,\penalty\@m}%
       \hbox{\csname b@\@citeb\endcsname}%
     \else%
      \advance\@tempcntb\@ne%
      \ifnum\@tempcntb=\@tempcntc%
      \else\advance\@tempcntb\m@ne\@citeo%
      \@tempcnta\@tempcntc\@tempcntb\@tempcntc\fi\fi}}\@citeo}{#1}}%
 
\def\@citeo{\ifnum\@tempcnta>\@tempcntb\else\@citea
  \def\@citea{,\penalty\@m}%
  \ifnum\@tempcnta=\@tempcntb\the\@tempcnta\else
   {\advance\@tempcnta\@ne\ifnum\@tempcnta=\@tempcntb \else
\def\@citea{--}\fi
    \advance\@tempcnta\m@ne\the\@tempcnta\@citea\the\@tempcntb}\fi\fi}

\catcode`\@=12

\begin{document}

\title{\bf  Gluon Saturation and the Formation Stage of Heavy Ion Collisions}
\author{Larry McLerran}
\maketitle
\begin{center}
 Physics Department and Riken Brookhaven Research Center, Building 510A\\ Brookhaven National Laboratory,
  Upton, NY-11973, USA
\end{center}

\begin{abstract}
 The high energy limit of QCD is controlled by very high energy density gluonic
 matter, the Color Glass Condensate.  In the first instants of the collisions of two sheets of Colored Glass Condensate, a Glasma is formed with longitudinal  flux tubes of  color electric and color magnetic fields.
 These flux tubes decay and might form a turbulent liquid that eventually thermalizes into a
 Quark Gluon Plasma.
\end{abstract}

\section{Introduction}

Gluons are ultimately responsible for  greatest part of the mass of  visible matter in the universe.
The gluonic contribution to mass  has its origin in the gluonic cloud which surronds light mass quarks.  This cloud results in about $99\%$ of the mass of a proton or neutron.

Gluons  generate the strong force.  This force permanently confines quarks and gluons inside of nucleons.  The nucleon-nucleon force, responsible for making atomic nuclei,
has its origins in gluons.  

We know very little of gluons by direct experiment.  Their existence is inferred indirectly.  
Approximately 1/2 of the momentum of a fast moving proton is in gluons.  This is known because the contribution of directly measured quarks only gives one half of the total momentum of the proton.  By measuring the distributions of quarks inside a proton and their variation
with the momentum scale of the measuring probe, one can extract  the implied distribution of gluons.  By  measurement of the production of high transverse momentum jets of particles produced in
collisions of strongly interacting particles, one can measure gluon distributions, but with strong restrictions that the momentum scale of the jets is large, and with uncertainties associated with jet 
hadronization.  
\begin{figure}[htbp]
\begin{center}
\includegraphics[width=0.60\textwidth]{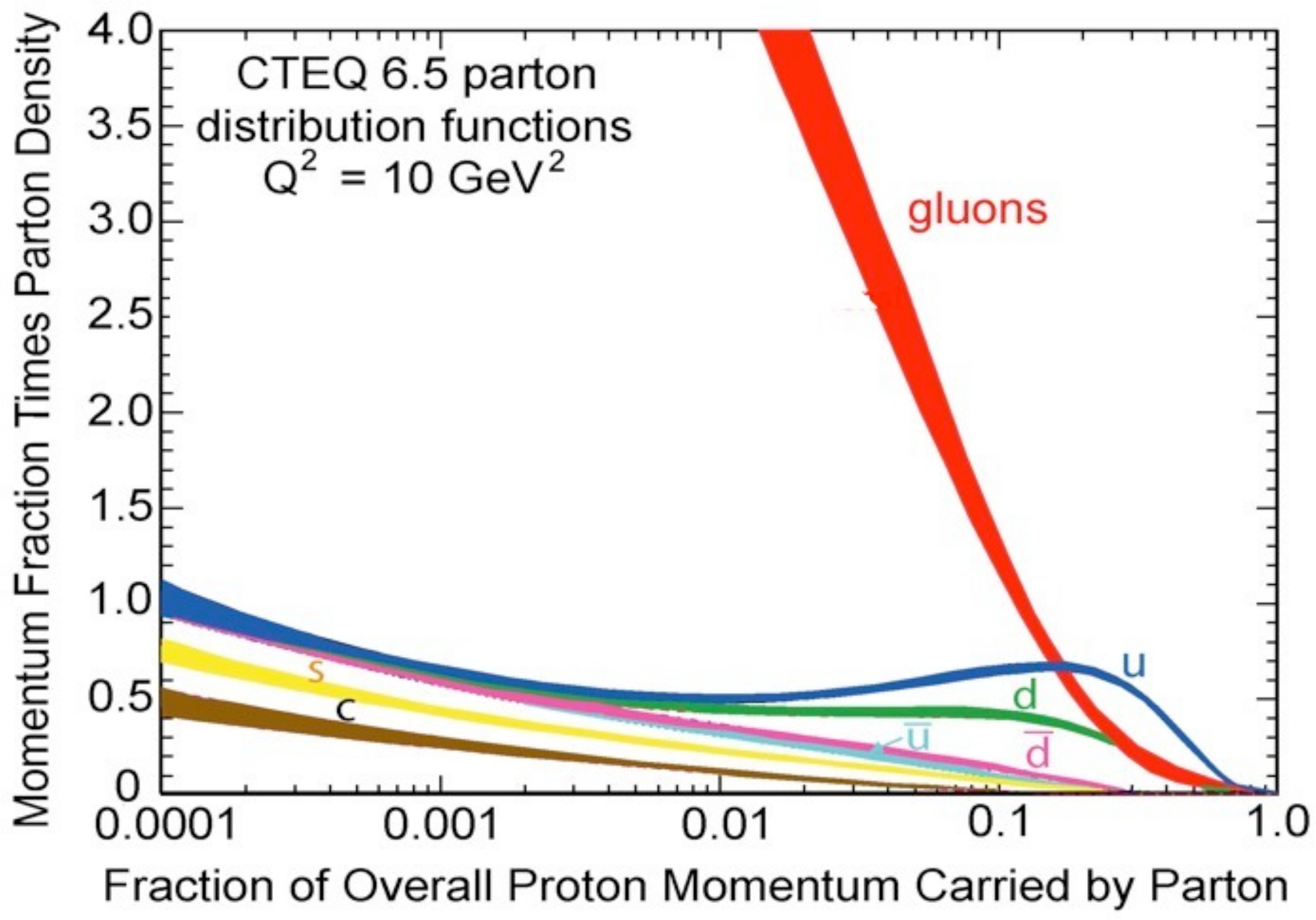} 
\end{center}
\caption{The quark and gluon distributions inside proton.
}
\label{distribution}
\end{figure}

The distribution of quarks and gluons inside a proton is shown in Fig. \ref{distribution}.\cite{hera}
The number distributions are plotted as a function of $x$ which is the fractional momentum of the 
constituents of a proton in a reference frame where the proton moves very fast.  
The typical momentum scale at which these gluons are probed is $Q^2 = 10~GeV^2$.
Note that below
$x \sim 0.1$, gluons dominate the proton wavefunction.

Experiments using very high momentum scales probe the short distance limit of QCD,
and such measurements provide most of our information about gluons. 
(For example, in the plot of Fig. \ref{distribution}, $Q^2 = 10~GeV^2$, which is much larger
than the QCD scale, $\Lambda_{QCD}^2 \sim 0.04~GeV^2$).
 Short distance measurements
are associated with only a tiny fraction of the processes which take place in high energy
scattering experiments.
Understanding typical properties of high energy collisions has been elusive.

The high energy limit is not the same as the short distance limit, since typical processes such as particle production do not
have momentum scales of order the beam energy.  Typical transverse momenta of produced particles grow quite slowly with energy.   Gluons 
control the cross sections for the collisions of high energy particles, and the total multiplicity of produced
particles.  This is because, as noted above, gluons dominate the hadronic wavefunction.  These gluons must be ultimately of central importance for the description of inelastic and quasi-elastic or diffractive collisions. 

The space-time description of high energy collisions is essential for any quantitative study.
Such a description was partially understood within the context of the parton model.\cite{bjfeynman}  The parton model however leaves many issues unresolved.  For example,
the cross section for the production of a particle with a typical average transverse momentum in a  high energy proton-nucleus collision scales like $A^{2/3}$, not A.  The probe nevertheless penetrates the entire nucleus.  Therefore, there must be coherence in the production process.  The parton model however
describes classical particles, and does not have this coherence.  Also, the total multiplicity of
particle production is not possible to compute within the parton model due to infrared divergences.  These
divergences are cured
when one allows for coherence. 

In collisions of heavy ions, one would like to have a complete description of the evolution of the matter
from the instant of collision.  In descriptions which invoke the parton model, this is only possible at some time later than a formation time $\tau_0 \sim 1~Fm/c$.   This restriction on time arises because before this time, coherent effects associated with the wavefunction of the nucleus are important.   At earlier times, much interesting physics occurs, such as the production of matter in the collisions, its interaction and possible  thermalization. 
 
There is now a QCD based theory of the high energy limit of QCD.\cite{mv}-\cite{iancucgc}  A surprising consequence of this theory is that it involves the properties of matter composed of gluons, that is the behavior of gluons in bulk, where the typical number of gluons is large
and the typical separation between gluons is much less than the size of the system.    This new form of matter generalizes our conception of QCD matter at finite temperature and density, the Quark Gluon Plasma (QGP).  
\begin{figure}[htbp]
\begin{center}
\includegraphics[width=0.30\textwidth]{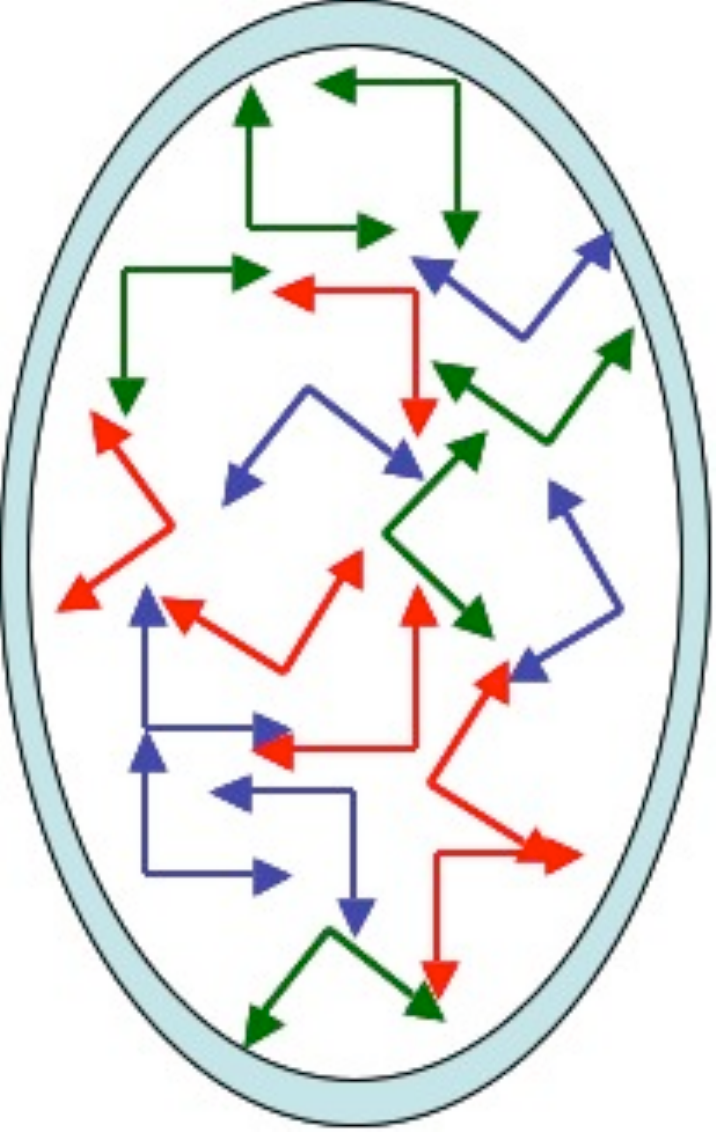} 
\end{center}
\caption{ A sheet of Colored Glass .
}
\label{onesheet}
\end{figure}

The Color Glass Condensate (CGC) is the matter which describes the gluons inside a strongly interacting particle when the typical longitudinal momentum fraction is small.  These are the wee gluons of the parton model.  It turns out that this matter is highly coherent and weakly coupled, although the interactions are enhanced due to coherence. (Coherence is also what makes the intrinsically weak gravitational force strong.  The intrinsic interaction between two protons is incredibly weak, but because all of the protons in large objects act together coherently, the resulting forces are strong.)

The CGC sits in a Lorentz contracted sheet of matter in a high energy hadron.  One can imagine a high energy hadron as a sheet of Colored Glass. For this high energy hadron, the color electric and magnetic fields are embedded in the sheet, since they originate and terminate on charges within each sheet,
and because a Lorentz boosted dipolar field becomes planar.  In fact these fields look like the
electromagnetic fields of a photon, with $\vec{E} \perp \vec{B} $, and correspond to Lorentz boosted Coulomb fields  The essential
difference between these fields and the Lorentz boosted Coulomb field of  charged particle is that
the polarization and color of the fields is a random variable, whose distribution will be determined by the theory of the Color Glass Condensate, and that the density of the gluon field is very high compared
to $\Lambda^2_{QCD}$.  The glassy sheet corresponding to a hadron is shown in Fig. \ref{onesheet}.
These features of the Color Glass Condensate will be motivated in later sections.
A single hadron can be probed by using a high energy electron.  This was the method of the HERA experimental program.\cite{hera}  
\begin{figure}[htbp]
\begin{center}
\begin{tabular}{l l l}
\includegraphics[width=0.45\textwidth]{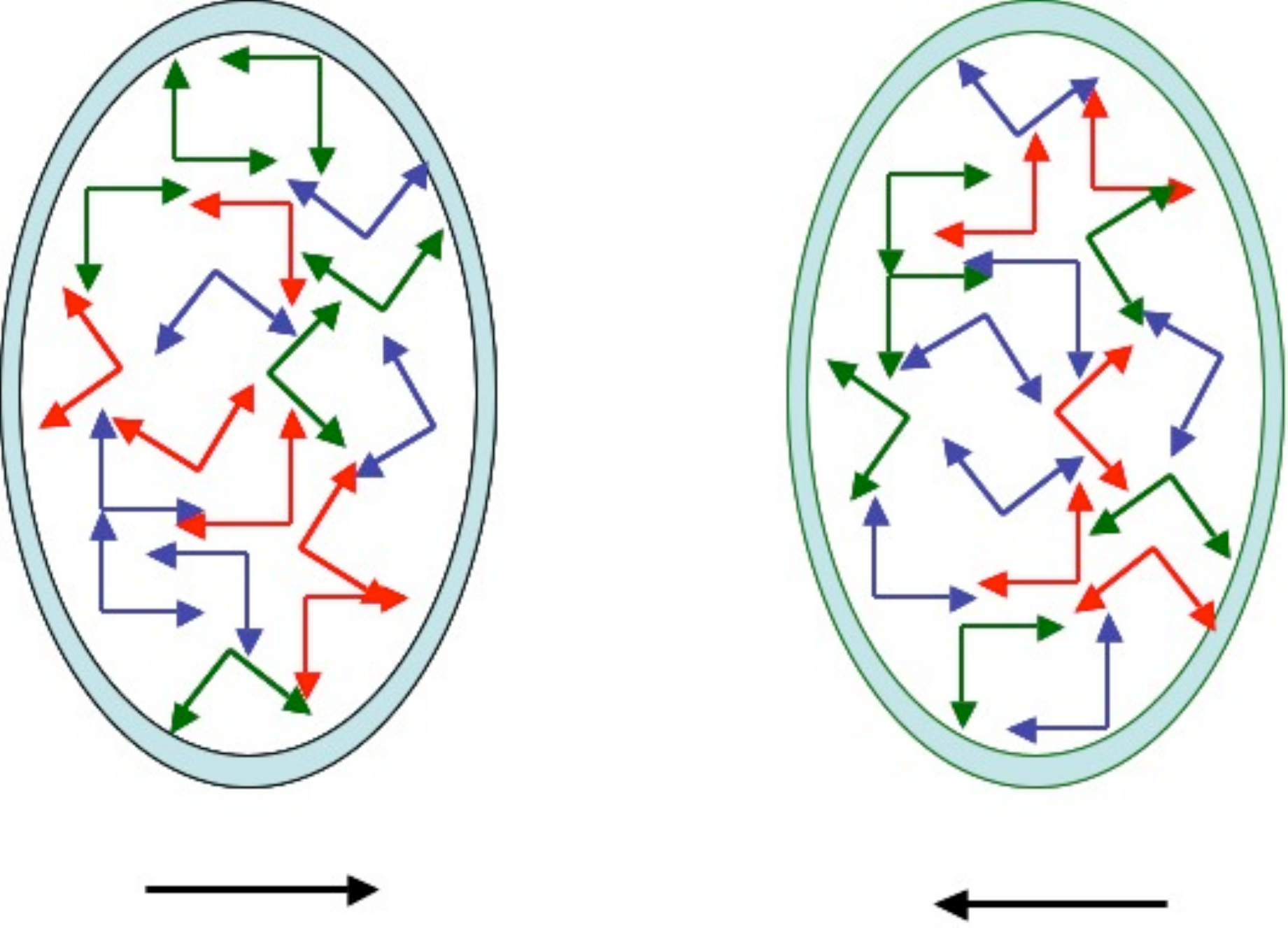} & & 
\includegraphics[width=0.45\textwidth]{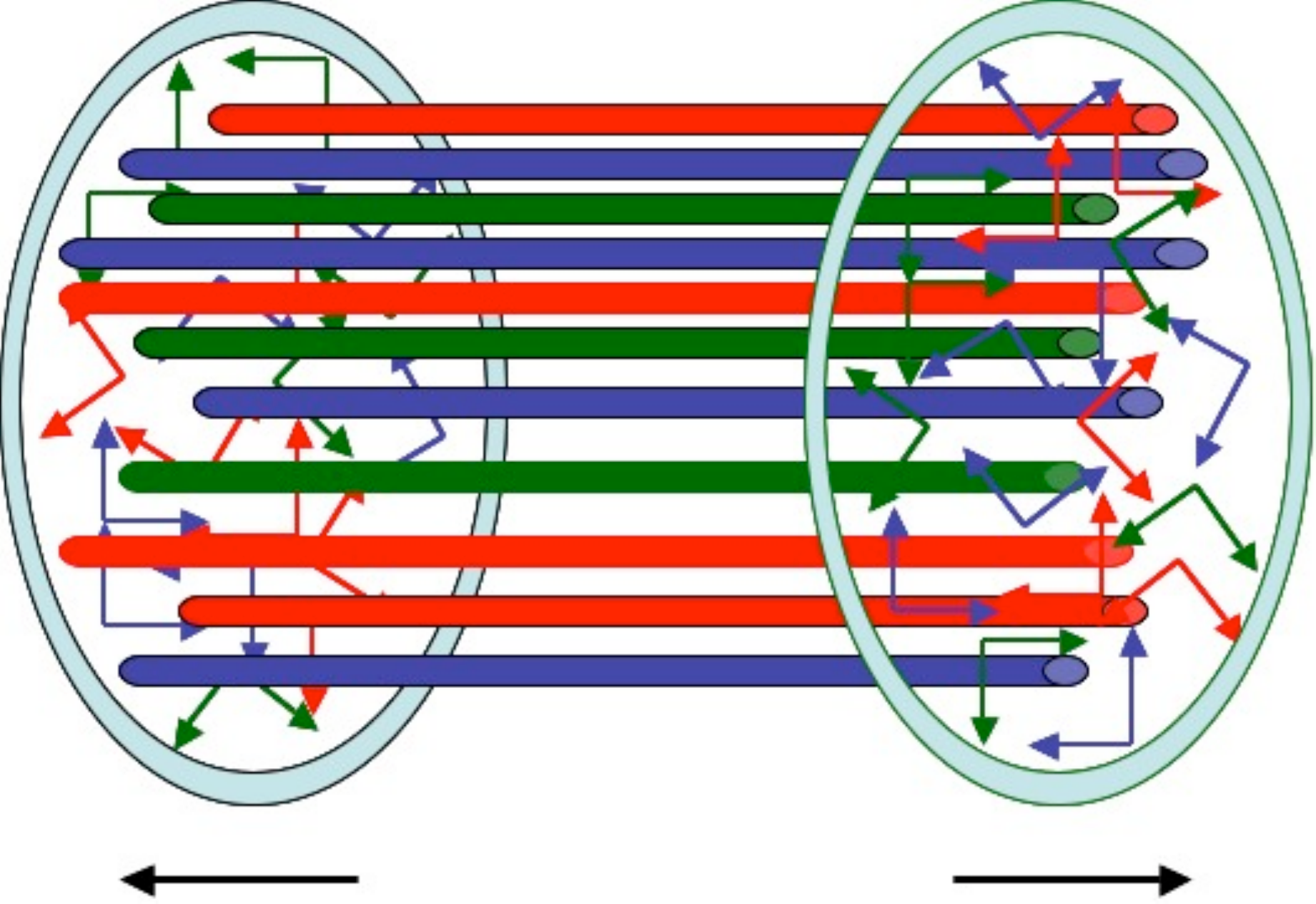}\\
 & & \\
 a & & b \\
\end{tabular}
\end{center}
\caption{(a) Two sheets of Colored Glass beginning to collide. (b) Two sheets of Colored Glass after their collision including the longitudinal flux  of color electric magnetic field generated by the collision.
The longitudinal fields are represented by the tubes on the figure.}
\label{sheetonsheet}
\end{figure}

A possible confusion concerning the CGC is the Lorentz invariance of the description.  We imagine
the hadron in a fast moving frame as a sheet of Color Glass.  It would seem we could boost to the rest frame, and all we should see are valence quarks.  The evasion of this paradox is that there are many very different states which contribute to the Fock space wavefunction of a hadron. For low energy processes, states with valence quarks and a few gluons dominate.  For high energy processes, the important states for typical processes have many gluons in them.  It is these many gluons states which comprise the CGC.

In high energy hadronic collisions, two sheets of Colored Glass collide.  These two sheets of Colored
Glass pass through one another, and in the short instant of time that this takes, the color electric and magnetic fields dramatically change.    Each sheet acquires an equal and opposite color electric and color magnetic charge density.  This results in strong longitudinal fields in the region between the sheets after they have passed through one another.  It is these longitudinal fields which are the initial conditions for the Glasma.\cite{weigert}-\cite{lappi}  Such a collision is represented graphically in Fig. \ref{sheetonsheet}.

As the Glasma evolves, the strong color fields decay and interact.  Expansion of the system results in a 
weakening of the field strength.  The fields eventually turn into weak radiation fields corresponding  to gluons.  In the process of decay, they  might thermalize through a variety of chaotic and turbulent processes.

In this paper, I will present the motivations for the CGC and Glasma, and an intuitive description of their properties.  I will discuss experimental evidence for these hypothetical forms of  matter, and how
further experimental study might verify their existence and test their properties.

\section{Motivation for the Color Glass Condensate}

The basic idea for the Color Glass Condensate is  a consequence of a few simple experimental observations.
The first is that the apparent size of a hadron varies very slowly with energy. This is known directly from experimental measurement of $pp$ and $p\overline p$ cross sections.  There are rigorous limits
on the rate of growth of cross sections at high energy.  The Froissart bound\cite{froissart} states that
\begin{equation}
		\sigma_{hadron} \le {A \over m_\pi^2} ln^2(E/E_0)
\end{equation}
where $A$ and $E_0$ are constants.  On the other hand, the HERA data on the gluon structure functions,\cite{hera} and theoretical arguments\cite{lipatov} give a growth in the number of gluons as
\begin{equation}
      N_{gluons} \sim B ~(E/E_0)^\delta
\end{equation}
where B and $\delta$ are constants and $\delta \sim 0.2 - 0.3$ at HERA energy.
This means that the gluon density per unit area has to grow.

The lowest energy gluon accessible for a hadron with energy $E$ is $x_{min} \sim \Lambda_{QCD}/E$.
The high energy limit is therefore the low $x$ limit.  It is also the limit of high gluon density.\cite{glr}-\cite{blaizot}
 
A density per unit area has the dimensions
of a momentum squared, so we introduce the concept of a saturation momentum,
\begin{equation}
    Q_{sat}^2 \sim N_{gluons}/\sigma_{hadron}
\end{equation}
where $\sigma_{hadron}$ is the high energy cross section at some energy $E$ and $N_{gluons}$ is the number of gluons at that energy.  This saturation momentum grows like a power of energy as the energy
increases, and at some point becomes much larger than $\Lambda^2_{QCD}$.  This means that
the gluons are so tightly packed inside a hadron  that their typical interaction  strength become weak,
$\alpha_{S}(Q_{sat}) << 1$.  We therefore come to the remarkable conclusion that the high energy limit
of QCD should be a weak coupling limit.   

Weak coupling limits of field theories like QCD allow for systematic approximations and computations.  The weak coupling limit is however not the limit of weakly interacting gluons.  This cannot be the case since
processes such as multi-particle production are divergent in simple perturbative expansions.  However, if there is a high density of gluons, such processes must be computable using weak coupling methods.

This apparent paradox can only be resolved if we allow for the gluon fields to be highly coherent.  This
situation is analogous to Quantum Electrodynamics, where the intrinsic interaction strength of fields is small, but one can have strong fields, as for example around a high Z atom.  For such an atom, the field strength is
characterized by $Z\alpha$, and when doing computations one must include effects to all orders in
$Z\alpha$, but there is a systematic expansion in powers of $\alpha$.  There is a well defined method of computation for processes in strong Coulomb fields.\cite{greiner}

For the high energy limit of QCD, the only quantity which can characterize strong fields is $1/\alpha_S$.  
If a field has an intrinsic strength of order $1/g$, then the typical occupation number of a gluonic state is
$<a^\dagger a > \sim 1/\alpha_S$.   This can be simply understood in terms of phase space density,
\begin{equation}
          {{dN} \over {dyd^2p_T d^2r_T}} = \rho
\end{equation}
The effective action in terms of $\rho$ should push $\rho$ away from $\rho = 0$ so as  to build up a condensate This means that for small $\rho$,
the energy becomes more negative as $\rho$ increases.  For sufficiently large $\rho$, the gluon interactions are repulsive, and the energy density must increase.  We therefore must minimize
\begin{equation}
                  E \sim -\kappa \rho +\kappa^\prime \alpha_S \rho^2
\end{equation}
 where $\kappa$ and $\kappa^\prime$ are two positive constants.  Extremizing gives,
$\rho \sim 1/\alpha_S$.

We now understand two of the words associated with the Color Glass Condensate.  Color comes
from the color of the gluons.  Condensate comes from the highly occupied bosonic states, somewhat analogous to the Bose condensate of superconductors and superfluids.      

The next ingredient is the idea of Glass.  This arises because if we are at a fixed high energy, the lowest 
$x$ gluons have as a source the color fields of the higher $x$ gluons.  The low $x$ gluons are the
Lorentz boosted Coulomb fields of the high $x$ gluons.  The classical field associated with these low
$x$ gluons evolve on a time scale characteristic of the fast gluons.  These fast gluons have their time scale Lorentz time dilated relative to a natural time scale such as $t \sim 1/Q_{sat}$.  Therefore, the slow gluons at low $x$ also evolve very slowly compared to a natural time scale.  Systems that evolve on time scales long compared to natural time scales, are generically glasses.  An ordinary silicate glass is a liquid on time scales of tens of thousand of years, which is very long compared to a time scale for evolution of an ordinary liquid.

The name Color Glass Condensate has several paradoxical features.  Most condensates have a single valued order parameter.  Here the condensate field is the gauge field.  It must be averaged over different local configurations in order to have a gauge invariant theory.   For the glass,
these configurations do not interfere with one another,  so there is a meaning to each configuration of the condensate, but this can only be expressed in terms of gauge invariant variables which measure
fluctuations.   One of the problems with naming an object such as the CGC is that it is conceptually new and has features analogous but,  at the same time,  distinctively different from known condensed matter systems.
		
The idea of slow gluons associated with a classical field and fast gluons associated with gluonic 
sources implies some a priori separation between the fast and slow gluons.  This would of course be frame dependent.  The beauty of the theory of the CGC is that it allows for this separation to be arbitrary.  
There is an effective theory for any separation of fast from slow, but the parameters of that theory depend
upon the separation scale.  The dependence of these parameters on the separation scale is determined
by renormalization group equations.
\begin{figure}[htbp]
\begin{center}
\includegraphics[width=0.40\textwidth]{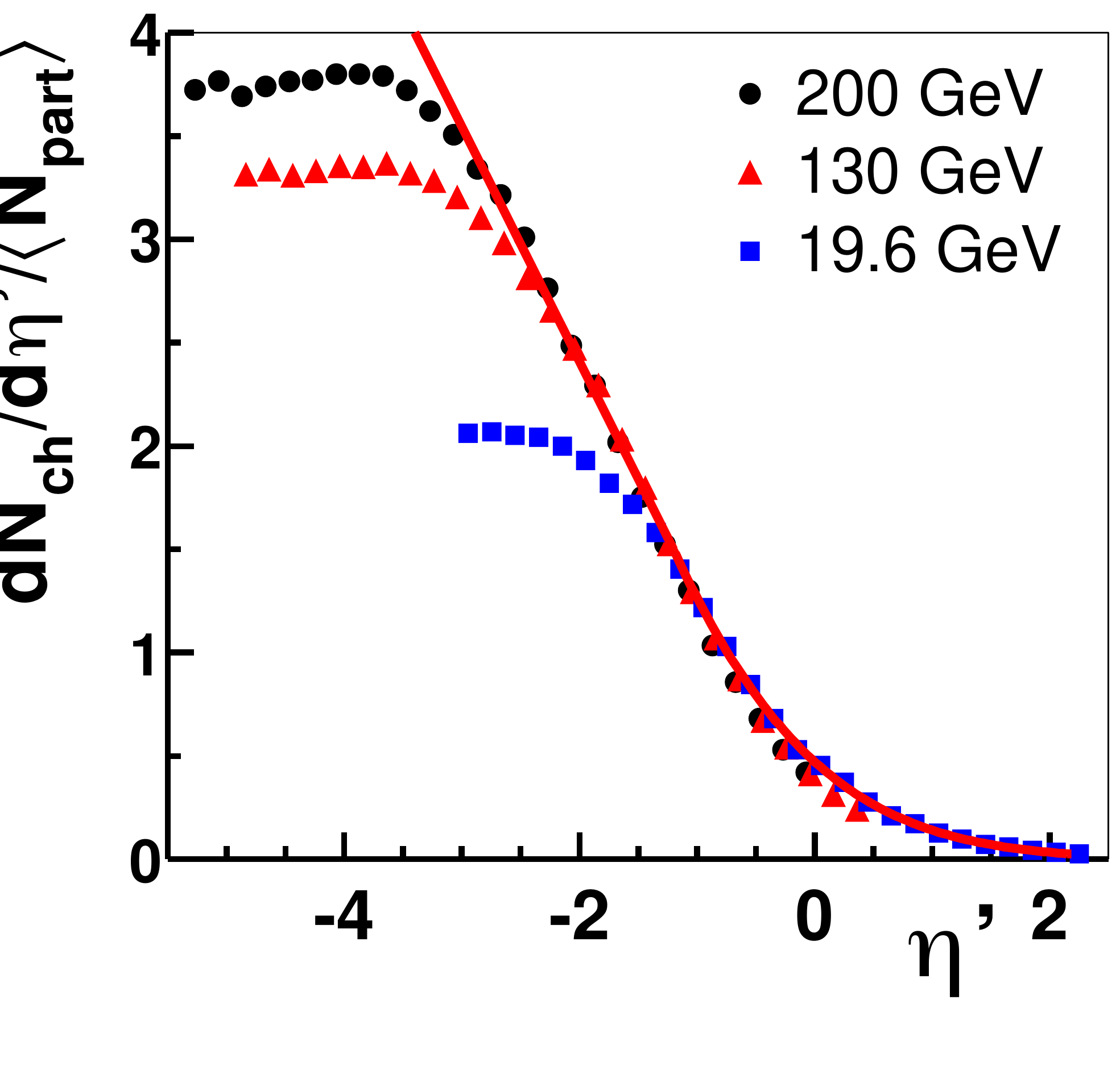} 
\end{center}
\caption{Limiting fragmentation as measured by Phobos in Au-Au collisions at RHIC .
}
\label{limfrag}
\end{figure}

The renormalization group description for the CGC can be motivated by the observation of limiting fragmentation.	This is shown in Fig. \ref{limfrag}.\cite{whitepaper}  Distributions of particles are plotted as a function
of rapidity, which is essentially the logarithm of the energy of the produced particle.  Results are for three different energies.  The distributions start at the rapidity the beam nucleons.  At low energies,
the curve rises and flattens out when one gets to the low $x$ region, which corresponds to rapiditites
much less than that of the beam nucleons.  At higher values of energy,
the fast moving particles trace the same curve, but break off at a lower value of $x$.    It is as if the high $x$ degrees of freedom are frozen, and the only effect of going to higher energies is to add in lower
$x$ degrees of freedom.  The density of these low $x$ degrees of freedom increase as we go to high energy, so whatever effective theory describes their properties must have parameters which change as the separation scale between slow (small x) and fast (large x) degrees of freedom change.

\section{Mathematics of the Color Glass Condensate}

Before proceeding to a theory of the CGC, it is useful to review a few kinematic approximations.  
I first introduce the light cone coordinates
\begin{eqnarray}
                p^\pm  & =  & (E \pm p_z)/\sqrt{2} \nonumber \\
                x^\pm & = & (t \pm z)/\sqrt{2}
\end{eqnarray}

For particles produced in high energy collisions,
the rapidity is 
\begin{equation}
y = {1 \over 2}~ ln\left\{{p^+\over p^-} \right\}
\end{equation}
The longitudinal momentum is $p_z$.  In deep inelastic scattering, the rapidity variable is
\begin{equation}
 y_{dis} =-ln(1/x)
 \end{equation}
 A little algebra shows that
 \begin{equation}
  y = -y_{dis} +ln(P^+_{proj}/M_T)
  \end{equation}
  where $M_T = \sqrt{p^+p^-}$.  These are the same except for a sign and the overall shift in rapidity.  Note however that  their application is quite different:  The rapidity of deep inelastic scattering is used to describe the constituents of a hadron, and the rapidity of high energy collisions is used for the description of produced particles. 
  
Space-time rapidity for particle scattering is
\begin{equation}
\eta = {1 \over 2} ~ln \left\{ {x^+ \over x^-} \right\}
 \end{equation}
 and for deep inelastic scattering, $\eta_{dis} =  ln(X^-_{proj}/x^-)$.  Clearly up to an overall sign and an overall constant, these rapidities are  the same up to uncertainty of order one unit of rapidity since the space-rapidities are simply related
 to the momentum space rapidities by the uncertainty principle relation,
 \begin{equation}
          x^\pm p^\mp \sim 1
\end{equation}          
          
The correlation between space-time and momentum space rapidities means that there is a correlation between the space-time and momentum space region of particles.  This lets us identify fast particles in momentum space with high rapidity particles in coordinate space.  

In the CGC, the fast particles are treated as sources,
\begin{equation}
            J^\mu_a(x) = \delta^{\mu +} \rho^a(x_T,\eta)  
\end{equation}
where $\eta$ is the space-time rapidity defined in Eq. 10..  There is no dependence upon the
light cone time $x^+$ because of the glassy nature of the matter.  If we want to compute a physical
quantity, we do a path integral with measure
\begin{equation}
           \int _{Y_0} [dA] [d\rho] ~e^{iS(A, \rho]} ~e^{-F[\rho]}
\end{equation}
In this equation, the action in the external current generated by $\rho$ is $S$.  The integration  over the sources is controlled by a Euclidean weight.  This is because the glassy nature of the matter does not generate quantum interference between configurations with different values of sources.  This prescription is similar to what is done with spin glasses.  The integration
over $\rho$ is for rapidities greater than some scale $Y_0$, and the fields are taken at rapidities less 
than $Y_0$.  The functional dependence of $F$ is determined by renormalization group equations,
and for many purposes may be treated as a Gaussian.

If one considers quantities with rapidity not too much less than $Y_0$, then the integration over
$A^\mu$ is determined by solving the classical field equations.  This is because the interaction strength is small.  So the physics is determined by a classical field that is then integrated over various sources with a weight function $F[\rho]$.  The fields become  strong because the strength of the sources at rapidities $y > Y_0$ is large, that is, there are many fast gluon sources.

The solution for the classical field equation in the presence of a light cone source is easily constructed,
\begin{equation}
   D_\mu F^{\mu \nu} = J^\nu
\end{equation}
Let $A^\mu$ have only a plus component.  Then
\begin{equation}
   -\nabla^2_T A^+ = \rho(x_T,\eta)
\end{equation}
The only non-vanishing component of $F^{\mu \nu}$ is $F^{i+}$, which can be composed in terms of 
$\vec{E}$ and $\vec{B}$.  These fields are perpendicular to the beam direction and perpendicular to one another.  They have a longitudinal extent  which is very small,  since $x^- \sim e^{-\eta}$, and the $\eta$
value for the source is chosen to be large. This gives the picture of the CGC shown in Fig. \ref{onesheet}

When we do the computation for the CGC, we work in light cone gauge.  This is the gauge where partons have a simple interpretation in terms of the 
Fock space component of a fast moving hadron wavefunction. When one transforms the field above to light cone gauge, the only non-vanishing field is
\begin{equation}
     A^i = {1 \over {ig}} U(x_T,\eta) \nabla^i U^\dagger(x_T,\eta)
\end{equation}     
where 
\begin{equation}
   U(x_T,\eta) = P exp\left\{ig \int^\infty_\eta d\eta^\prime {1 \over {-\nabla_T^2}} \rho  \right\}
\end{equation}
For values of $\eta$ far away from the region of the extent of the source, the vector potential is
essentially a step function, being zero for $x^0$ negative, and non-zero for positive values

At the classical level, the fields $F^{\mu \nu}$ have an extent of the order the source size.  Nevertheless,
it is the extent of the vector potential which determines where the Fock space components of the 
nuclear wavefunction reside.  Fourier transforming in $x^0$ gives a wide distribution of momentum.
Of course, the parton description in terms of incoherent particles must break down for such fields,
since the electric and magnetic field are essentially at short longitudinal distance scales and
the vector potential is longer range.  This is a consequence of the coherence of the field.

If the rapidity of the quantity we wish to compute is too far different from $Y_0$, quantum corrections
of order $\alpha_S \Delta y $ can become large.  This determines the size of  the region over which the effective CGC theory holds,
\begin{equation}
  \Delta y << 1/\alpha_S
\end{equation}
The thickness of the source distribution follows from this equation as 
\begin{equation}
  \Delta z \sim {1 \over {Q_{sat}}} e^{-\kappa/\alpha_S}
\end{equation}
where $\kappa$ is some constant.   This justifies thinking of the sources as being on a thin sheet.

 In order to compute at rapidities far from $Y_0$, one must systematically integrate out the quantum fluctuations.  This can be done using renormalization group techniques and involves a loop computation in an arbitrary CGC external field.  The computation can be done with an explicit result.  The resulting equation determines the form of $F$.  The equation is called the JIMWLK equation and is of the form,
 \begin{equation}
             {d \over {dy}} e^{-F} = - H e^{-F}
 \end{equation}
The Hamiltonian $H$, is second order in derivatives of $\rho$  and all orders in $\rho$.\cite{kovner}-\cite{iancu}    (The mean field approximation to 
this equation is the Balitsky-Kovchegov equation.\cite{kovchegov1}-\cite{balitsky}).  There is no potential in the function space of $\rho$ for this Hamiltonian.  It is all kinetic energy, with a kinetic energy operator non-linear in $\rho$.  A Hamiltonian without a  potential describes diffusion. This means that the typical distribution in $\rho$ grows forever as $y \rightarrow \infty$.  This means ultimately implies the saturation momentum grows without bound in the infinite energy limit.  
\begin{figure}[htbp]
\begin{center}
\includegraphics[width=0.60\textwidth]{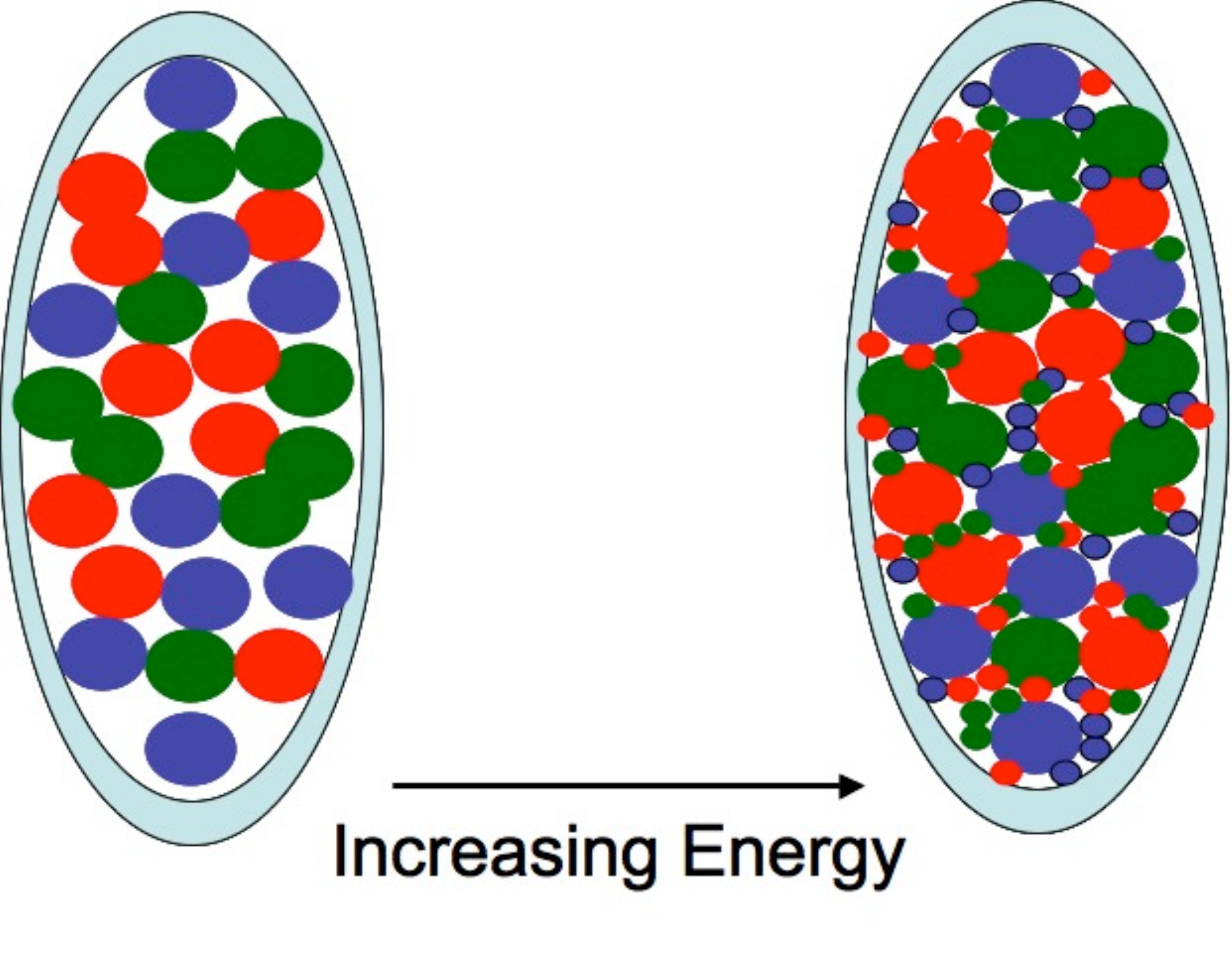} 
\end{center}
\caption{Saturation as a function of increasing gluon number corresponding to increasing energy.
}
\label{saturation}
\end{figure}

How can the number of gluons grow forever?  Surely for gluons of a fixed size, at some energy there
is no more space to pack them inside a hadron. 
To see how the growth in the number of gluons  might occur, we consider a simple model of saturation where gluons are considered as hard disks.  The gluons have an interaction strength of order $\alpha_S$, so that $1/\alpha_S$ gluons act like a hard disk.  The transverse size of a gluon corresponds to its De Broglie wavelength in the transverse plane, $r_T \sim 1/p_T$.  If we have gluons of some fixed size, we can pack them together until they overlap as shown in the left hand side of Fig. \ref{saturation}.  This means the number of gluons per unit area which are highly packed together is $Q_{sat}^2 \sim p_T^2/\alpha_S$  Now if we go to higher energy, we can add in more gluons of smaller size in the holes between the saturated gluons
of larger size.  This means that the saturation momentum can grow as the energy increases, and the
typical transverse momentum of the gluons also increases.  

{\it Saturation does not mean that the number of gluons or saturation momentum stops growing.}  It means that if we measure the number of gluons of some fixed transverse momentum scale, then at some energy,
the number of these gluons stops growing rapidly.  The growth continues for gluons of higher transverse momentum corresponding to smaller sizes.  This is a consequence of the diffusive nature of the evolution of gluon evolution
embodied in the JIMWLK equations.

\section{Phenomenological Consequences of the Color Glass Condensate} 

The CGC provides a first principles QCD description of a high energy hadron.  It may also be applied to the description of very high energy hadron-hadron collisions as will be presented in  a later section on the Glasma.  As such, it should explain generic properties of high energy lepton-hadron scattering
scattering non-diffractive and diffractive.  In this section, I present  some of the consequences of the CGC for lepton-hadron scattering.  I also outline some generic features of hadron-hadron scattering which do not require a detailed understanding of the Glasma,  that provides a theory of  the matter produced in hadron-hadron collisions.

A simple consequence of the CGC is that it provides a local theory of the matter inside a hadron.  There is only one dimensionful scale which characterizes this theory, the saturation momentum,  up to a weak dependence on $\Lambda_{QCD}$ which is implicit in the coupling constant dependence of the theory.  Energy dependence of physical quantities formulated within the CGC can only occur due to the
dependence of energy upon the saturation momentum.  This  leads to scaling relations
for various observables.
\begin{figure}[htbp]
\begin{center}
\includegraphics[width=0.50\textwidth]{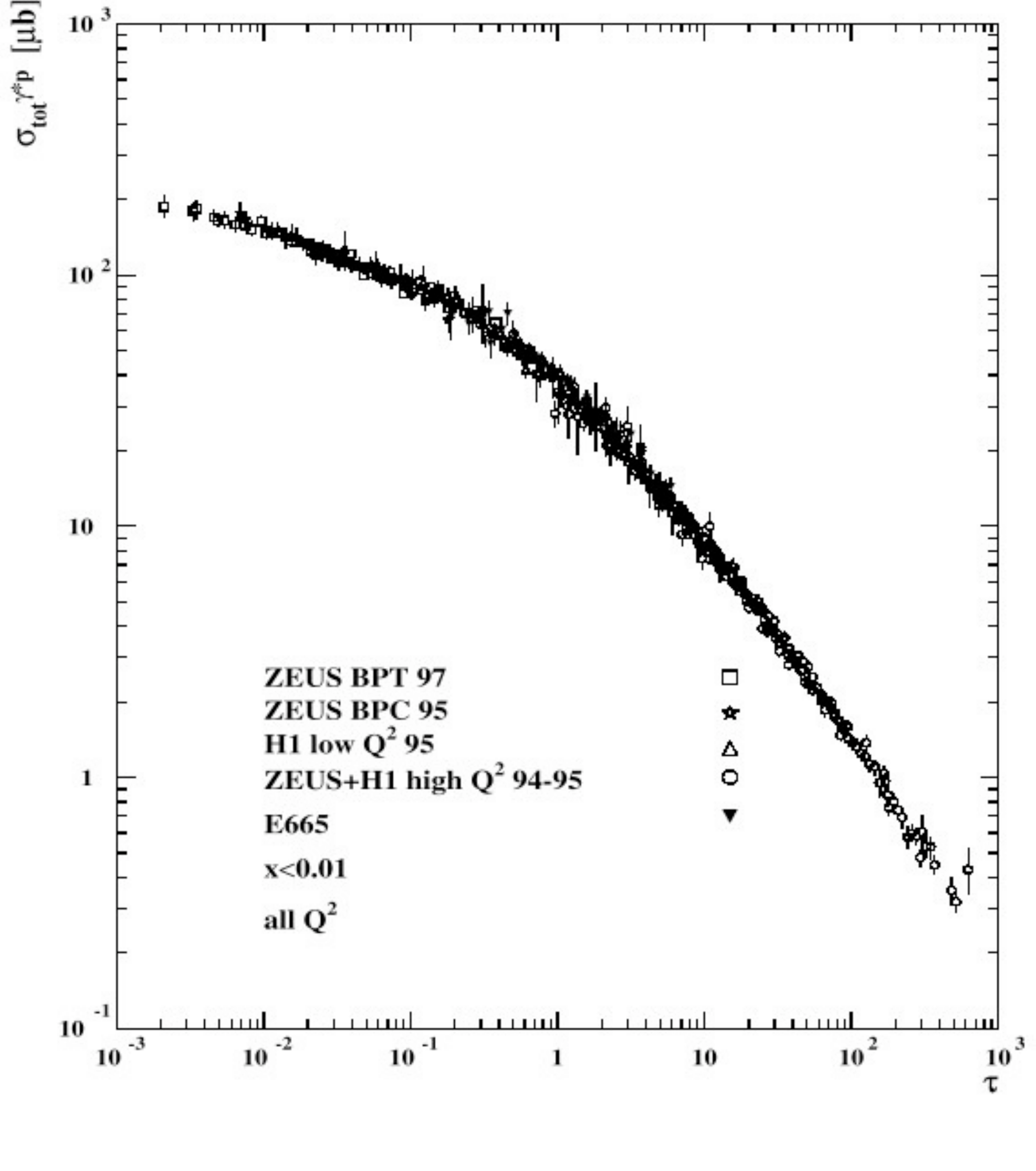} 
\end{center}
\caption{Geometric Scaling of the HERA data on deep inelastic scattering for $x \le 10^{-2}$.
}
\label{geomscaling}
\end{figure}

For example, the deep inelastic total cross section for virtual photon with momentum $Q$ should have the functional form,\cite{stasto}
\begin{equation}
  \sigma_{\gamma^* p} = F(Q^2/Q_{sat}^2)
\end{equation}
As shown in Fig. \ref{geomscaling}, this scaling is well satisfied in the HERA data for $x \le 10^{-2}$.
The $x$ dependence of the saturation momentum can be extracted from this comparison,
$Q_{sat}^2 \sim Q_0^2 (x_0/x)^\delta$ where $\delta \sim 0.2 - 0.3$. 

\begin{figure}[htbp]
\begin{center}
\includegraphics[width=0.45\textwidth]{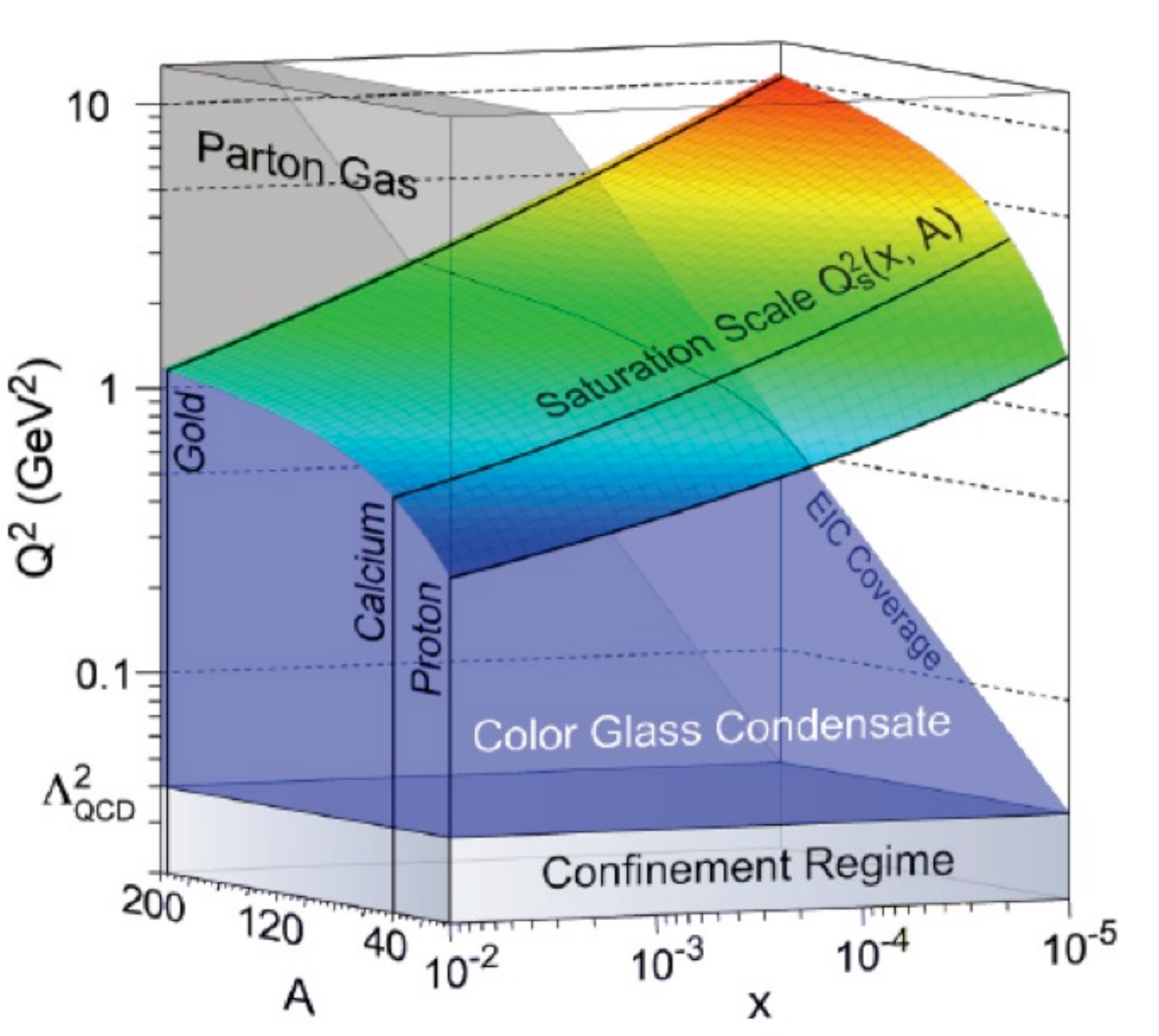} 
\end{center}
\caption{The saturation momentum as a function of $x$ and $A$. 
}
\label{satmomentum}
\end{figure}

One could make the saturation momentum higher by using nuclei.  We expect that the total number of gluons inside a hadron scales like $A$, the nucleon number.  This means the saturation momentum, which is proportional to the number of gluons per unit area should scale as $Q_{sat}^2 \sim A^{1/3}$.  Given the slow variation of the saturation  with $x$, it has been suggested that an electron-ion collider might probe the high density region of the CGC.\cite{erhic}  A plot of expected values of the saturation momentum for various $A$ and  $x$ is shown in Fig. \ref{satmomentum}.
  
The scaling behavior seen on the Fig. \ref{geomscaling}, includes values of $Q^2 >> Q_{sat}^2$.
this is surprising since the theory of the CGC should apply for $Q^2 \le Q_{sat}^2$.  In fact, even at large $Q^2 \le Q_{sat}^4 /\Lambda^2_{QCD}$, geometric scaling can be shown to work.\cite{kazu}   More generally, one can fit the quark and gluon distribution functions in theories which include saturation 
and one achieves a good and natural description of the low x data.\cite{gb}-\cite{munier}
  
The saturation momentum itself can be computed using the renormalization group equations.\cite{kazu}-\cite{dion}   Such a computation gives a saturation momentum which agrees with the data on gluon distribution functions from HERA.  Such computations determine  the effects of leading
twist shadowing from nuclear targets.

The CGC provides not only a description of deep inelastic structure functions, but also diffractive structure functions of the proton.\cite{gb}, \cite{kovchegov3}-\cite{kowalski}   Again, the agreement between phenomenological descriptions based on the CGC and the HERA data is quite good.

The CGC gives a heuristic understanding of the Froissart bound.\cite{kovner1}-\cite{ferreiro}  Assume that the distribution of gluons in a hadron factorizes into an impact parameter piece and a piece proportional to the rapidity dependence of the number of gluons.  We will assume that the impact parameter piece falls off exponentially at large distance with $2m_\pi$.  The total cross section is controlled by isosinglet exchange and the lowest mass isosinglet contribution is two pions.  The number distribution of gluons is therefore of the form
\begin{equation}
    {{dN} \over {d^2r_T}} \sim e^{-2m_\pi r_T} e^{\delta Y} 
\end{equation}
where $Y$ is the rapidity of the hadron, $Y \sim ln(E/E_0)$  The number $\delta$ comes from
the $x$ dependence of the gluon distribution function.  If we probe the hadron with an object of a fixed size, then the edge of the hadron is determined by penetrating a fixed number of gluons at some impact parameter $r_T$.  This means that
\begin{equation}
{{dN} \over {d^2r_T}} = constant
\end{equation}
or
\begin{equation}
  \sigma \sim R_T^2 \sim {{\delta^2 Y^2} \over {2\m_\pi^2}}
\end{equation}
which saturates the Froissart bound.

The CGC has also led to understanding of  shadowing in high energy $pA$ scattering.\cite{kovner3}-\cite{jamal}   In high energy $pA$ collisions, a projectile proton would scatter multiple times as it traverses a nucleus.  This would lead to a broadening of the momentum distribution.  On the other hand,
the evolution of the gluon distribution function of a nucleus to small x is cutoff by the saturation momentum which is larger for a nucleus than for a proton.  The larger cutoff in nuclei than in a proton implies that less gluons than would be naively expected are induced at small x.  Therefore, one might expect some shadowing which would reduce the number of gluons relative to the naive expectation that the gluon distribution function scales as $A$.   In more central collisions, both effects are enhanced.  These two effects for intermediate transverse momenta have opposite signs. It turns out that the first effect dominates at intermediate $x$ values for a nucleus, and the latter at small $x$ values.  The effect turns on quite rapidly as a function of $x$.  At $x$ values appropriate for SPS energies, the multiple scattering term dominates. At the $x$ values appropriate for the forward region (deuteron fragmentation region)of  $dA$ collisions, small $x$ values of the nuclear wavefunction are important, and one expects shadowing at intermediate $p_T$.  It is surprising that these effects roughly cancel in the central region of
$dA$ collisions at RHIC.\cite{whitepaper}  Both of these effects are included in the CGC description of the nuclear wavefunction, and quantitative comparisons have been done with the RHIC data with reasonable agreement.\cite{brahms}

\section{The Glasma and High Energy Nuclear Scattering}

\begin{figure}[htbp]
\begin{center}
\includegraphics[width=0.50\textwidth]{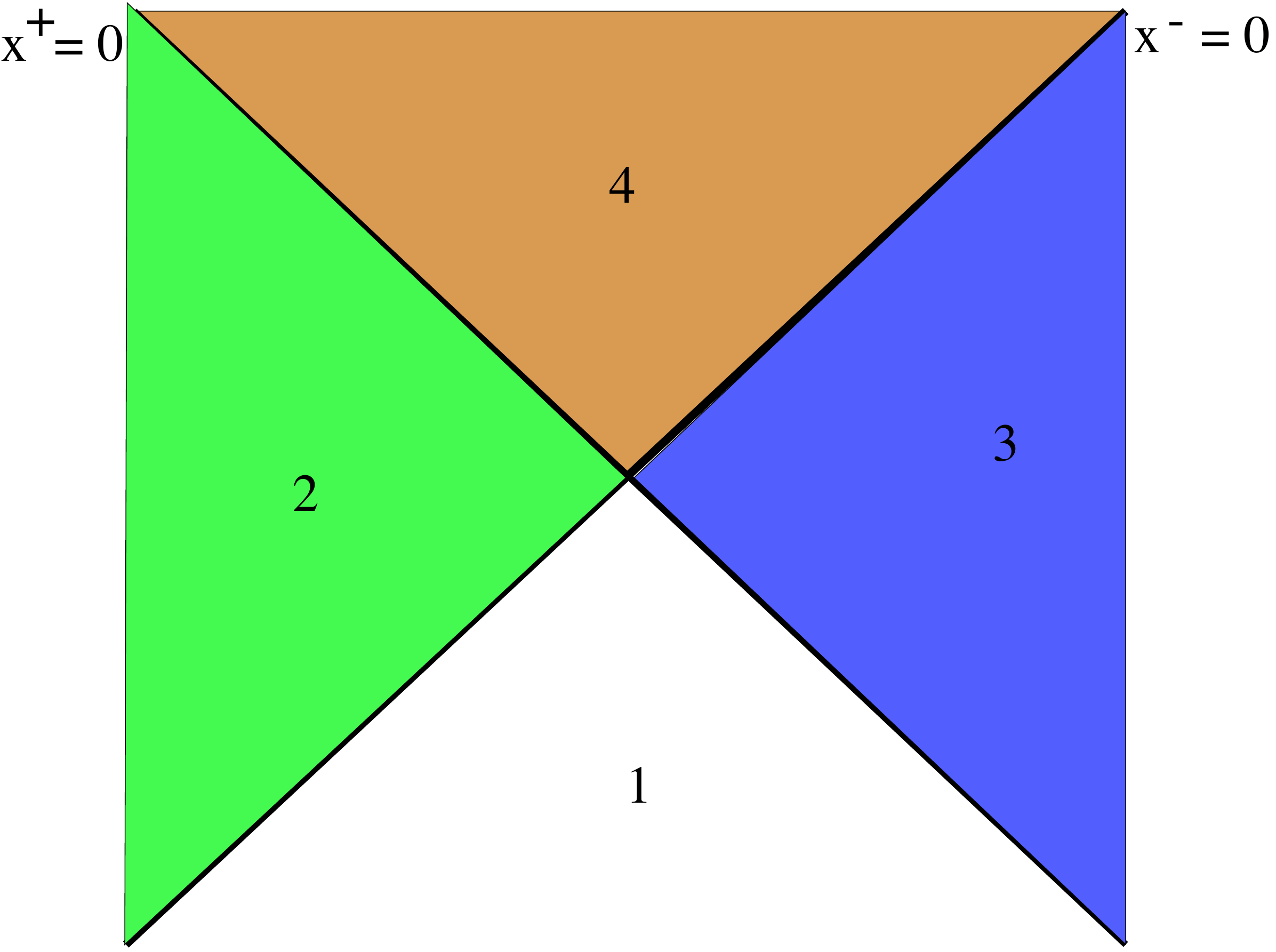} 
\end{center}
\caption{A lightcone diagram showing the various regions important for high energy hadron-hadron scattering.
}
\label{lightcone}
\end{figure}

\begin{figure}[htbp]
\begin{center}
\includegraphics[width=0.70\textwidth]{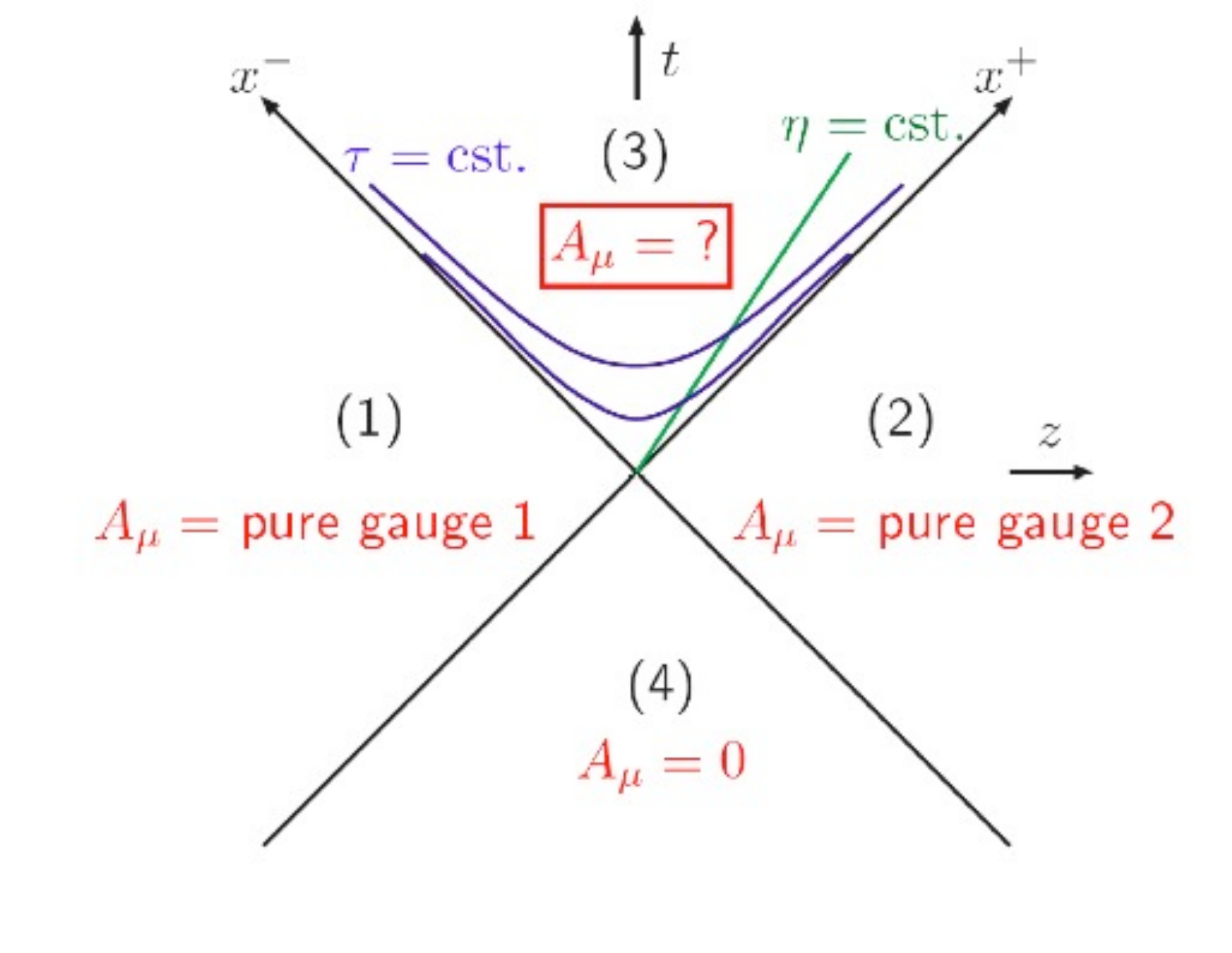} 
\end{center}
\caption{The fields associated with the collision of two high energy hadrons.
}
\label{spacetb}
\end{figure}

The collision of two sheets of Color Glass Condensate should describe hadron collisions at very high energy.  We first imagine that for negative time, there are two sheets of CGC flying towards one another at near light speed.  The initial nuclei are thought of as sources at $x^- = 0$ and $x^+ = 0$
The collision may be visualized in the lightcone diagram shown in Fig. \ref{lightcone}.
In region 1, corresponding  to the region behind both of the nuclei before the collision,  we set the vector potential to zero.
As we proceed from region 1 into region 2, we cross the sources where the nucleus with $x^- = 0$ sits.
The vector potential in this region, $A^i_1$ becomes non-zero so that the non-Abelian Maxwell equations can be satisfied.  Similarly in region 3, a different nonzero vector potential, $A^i_2$ is induced.  We might naively think we could solve the Maxwell equations in the forward light cone by setting 
$A = A^i_1 + A^i_2$ in region 4.  This can indeed generate the correct sources for the nuclei.  However, fields of this type will not solve the equations in the region 4 away from the light cone.  This is because the sum of two gauge transformations is not a gauge transformation for a non-Abelian theory.  The full solution in region 4 is time dependent, and involves more components of the vector potential.  There are non-zero field strengths $F^{\mu \nu}$ in the forward light cone.  The fields are represented in Fig. \ref{spacetb}.

Nevertheless, for times infinitesimal after the collisions has taken place, the vector potentials are indeed
given by $A^i = A^i_1 + A^i_2$.  The new feature which occurs immediately after the collision is that the
vector potential associated with hadron 1 can multiply the source strength of that of hadron 2,
and vice versa.  By the non-Abelian Maxwell equations, this means that a source of color magnetic and
color electric charge is induced on the two hadrons,
\begin{eqnarray}
   \nabla \cdot E_{1,2}^a & =  & - f^{abc} A^{b}_{2,1} \cdot E^c_{1,2} \nonumber \\
    \nabla \cdot B_{1,2}^a & =  & - f^{abc} A^{b}_{2,1}\cdot   B^c_{1,2}
\end{eqnarray}
One can verify that the charge densities on the two hadrons are equal and opposite.  There has to be
both color magnetic and color electric charge induced since the electric and magnetic fields in the CGC play a symmetric role, and the QCD equations are self dual under the interchange of $E$ with $B$.

The non-Abelian Maxwell equations therefore imply instantaneous charging of the sheets of Color Glass as they pass through one another.  This means that longitudinal electric and magnetic fields are instantaneously set up.  These fields are random on a transverse scale of order $1/Q_{sat}$.  This process appears to be instantaneous in time due to approximation that the sheets of CGC are infinitesimally thin.  In a previous section we argued that the true extent of the sheets of CGC are of order $e^{-\kappa/\alpha_S}/Q_{sat}$ thick, so that the time it takes the sheets to pass through one another is small but finite..  This is much smaller than the extent of the region in which these Glasma fields are produced.  

The reason  for the new name Glasma is because during the collision, the fields have very rapidly changed their topology from  transverse to   longitudinal.  Glasma is a hybrid word which is meant to convey that this new matter arises from the Color Glass Condensate and evolves to the Quark Gluon Plasma.

How should we imagine this spontaneous charging of the sheets of Color Glass?  When a pair of particles are produced in the central region, they need to have arisen from collisions of particles associated with the sheets of Color Glass.  Charge must be conserved however, so if charge is transferred to the produced particles,  an opposite charge must be induced in the sheets of Color Glass.

These flux tubes carry both electric and magnetic field.  They induce a large Chern-Simons charge
density, which may ultimately be responsible for effects associated with Chiral Symmetry restoration.
The anomalous $U(1)$ axial vector current is
\begin{equation}
  \partial_\mu J^\mu_5 = \kappa~ \partial_\mu K^\mu_{Chern-Simons} = \kappa ~\vec{E} \cdot \vec{B}
\end{equation}  
Here, $\kappa$ is a constant.

The total Chern-Simons charge can be computed for field configurations associated with the Glasma, and although the density is large, the integral vanishes. Nevertheless in the evolution of the classical field equations, Chern-Simons number may be generated\cite{rajucp}.  The generation of a net Chern-Simon number violates P and CP, and could be observed as fluctuations in high energy heavy ion collisions.\cite{harmen}  

The Glasma fields evolve with time and weaken due to expansion of the matter produced in high energy collisions.  The total number of gluons can be computed by directly solving these equations.  This number computed in perturbation theory would diverge because of the $1/p_T^4$ singularity of perturbation theory.  In the Glasma, the computation for $p_T >> Q_{sat}$ is 
\begin{equation}
        {1 \over {\pi R^2}} {{dN} \over {dyd^2p_T}} \sim {1 \over \alpha_S} {{Q_{sat}^4 } \over p_T^4}
\end{equation}
The singularity at small momentum arises incoherently resolving  individual charges on scales less than the inverse saturation momentum.   In fact, on scales larger than this, the color charges of each hadron appear
as multipoles with an average zero charge.  This reduces the potential associated with each hadron by two powers of $r_T$ at large $r_T$.  The multiplicity is therefore  no longer power law divergent at small $p_T$, and
\begin{equation}
       {1 \over {\pi R^2}} {{dN} \over {dyd^2p_T}} \sim {1 \over \alpha_S} 
\end{equation}
up to powers of $log(Q_{sat}/p_T)$.  This yields a finite result for the total multiplicity.

On dimensional grounds, the total multiplicity is
\begin{equation}
       {{dN} \over {dy}} \sim {1 \over \alpha_S} \pi R^2 Q_{sat}^2
\end{equation}
The factor of $1/\alpha_S$ is required because the fields which produce the CGC are classical, and contain of order $1/\alpha_S$ quanta.  Another way to understand this formula is that $1/Q_{sat}^2$ is the area of a color electromagnetic flux tube.  This tube decays into $1/\alpha_S$ particles per unit rapidity.  The formula is simply
\begin{eqnarray}
          {{dN} \over {dy}} &  =  & {1 \over \alpha_S} {{Area~ of~ hadron} \over {Area~ of~ flux~ tube}} \nonumber \\  & = &  (Gluons~ per~ flux~ tube) \times (Number~ of~ flux ~tubes)
\end{eqnarray}

Kharzeev and Nardi compared such a CGC inspired formula to the RHIC data, and predicted the dependence of the multiplicity upon the centrality of the collision.  There is an assumption that the number of gluons is the same as the number of pions, an assumption which would be true if there is instant thermalization and isentropic expansion.  Krasnitz and Venugopalan computed the non-perturbative constant in the relationship above.\cite{venugopalan}   

\begin{figure}[htbp]
\begin{center}
\includegraphics[width=0.90\textwidth]{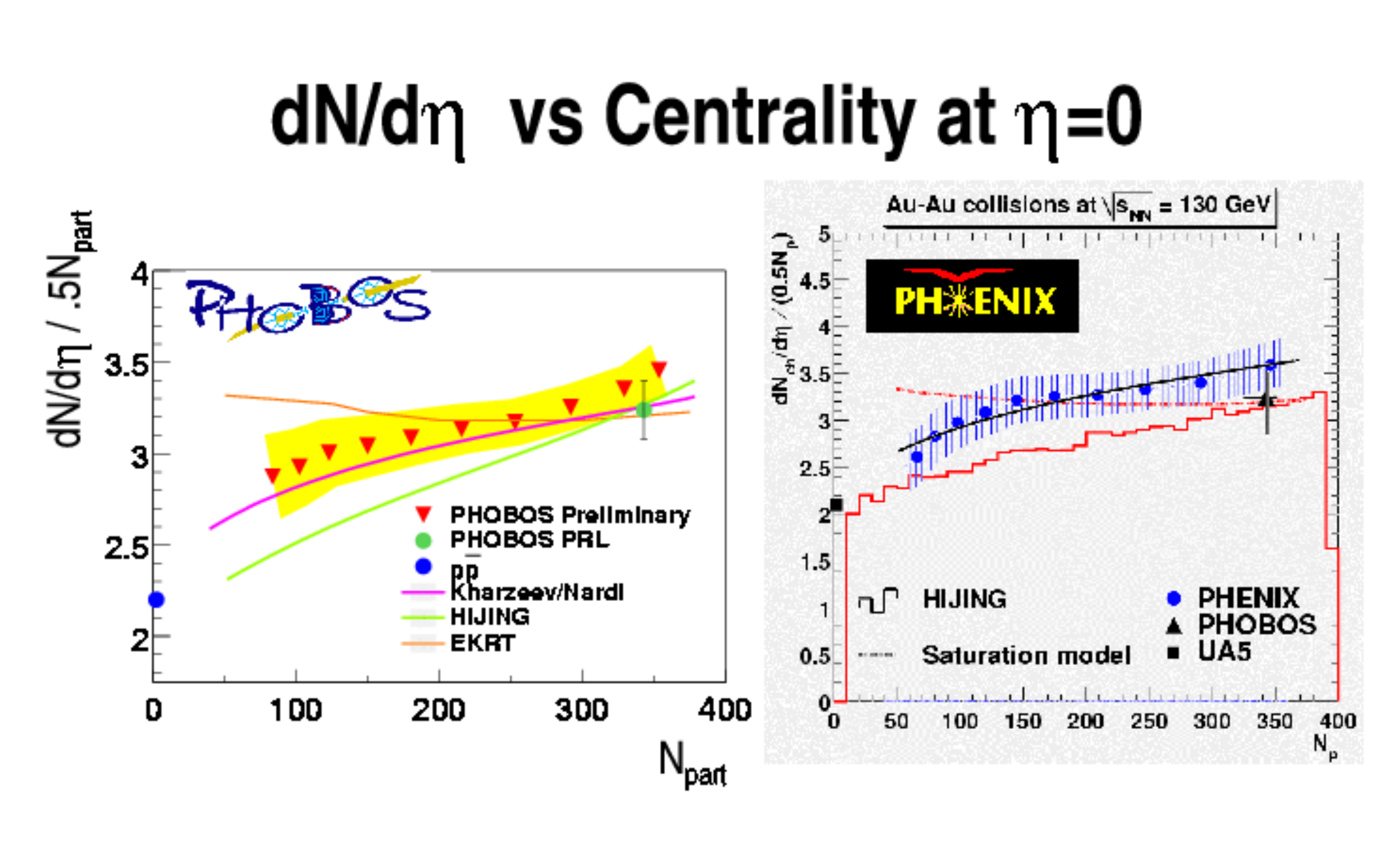} 
\end{center}
\caption{The dependence of the mutliplicity on centrality.  The solid line which goes through the data point with error bars is the Kharzeev Nardi prediction. 
}
\label{kn}
\end{figure}

The result of such a comparison is shown in Fig. \ref{kn}  What is plotted is the 
multiplicity scaled by the number of participants in the collision.  The saturation momentum squared
scales as $N_{part}^{1/3}$ and $R^2 \sim N_{part}^{2/3}$, so that the CGC predicts
\begin{equation}
            { 1 \over N_{part}} {{dN} \over {dy}} \sim {1 \over \alpha_S} 
\end{equation}
and $\alpha_S$ is slowly varying on account of its dependence on the running coupling constant.
The value of $Q_S$ that describes the RHIC multiplicities $Q_s \sim 1.0_1.2~GeV$
is consistent with the HERA and NMC data.\cite{lappihera}

\begin{figure}[htbp]
\begin{center}
\includegraphics[width=0.70\textwidth]{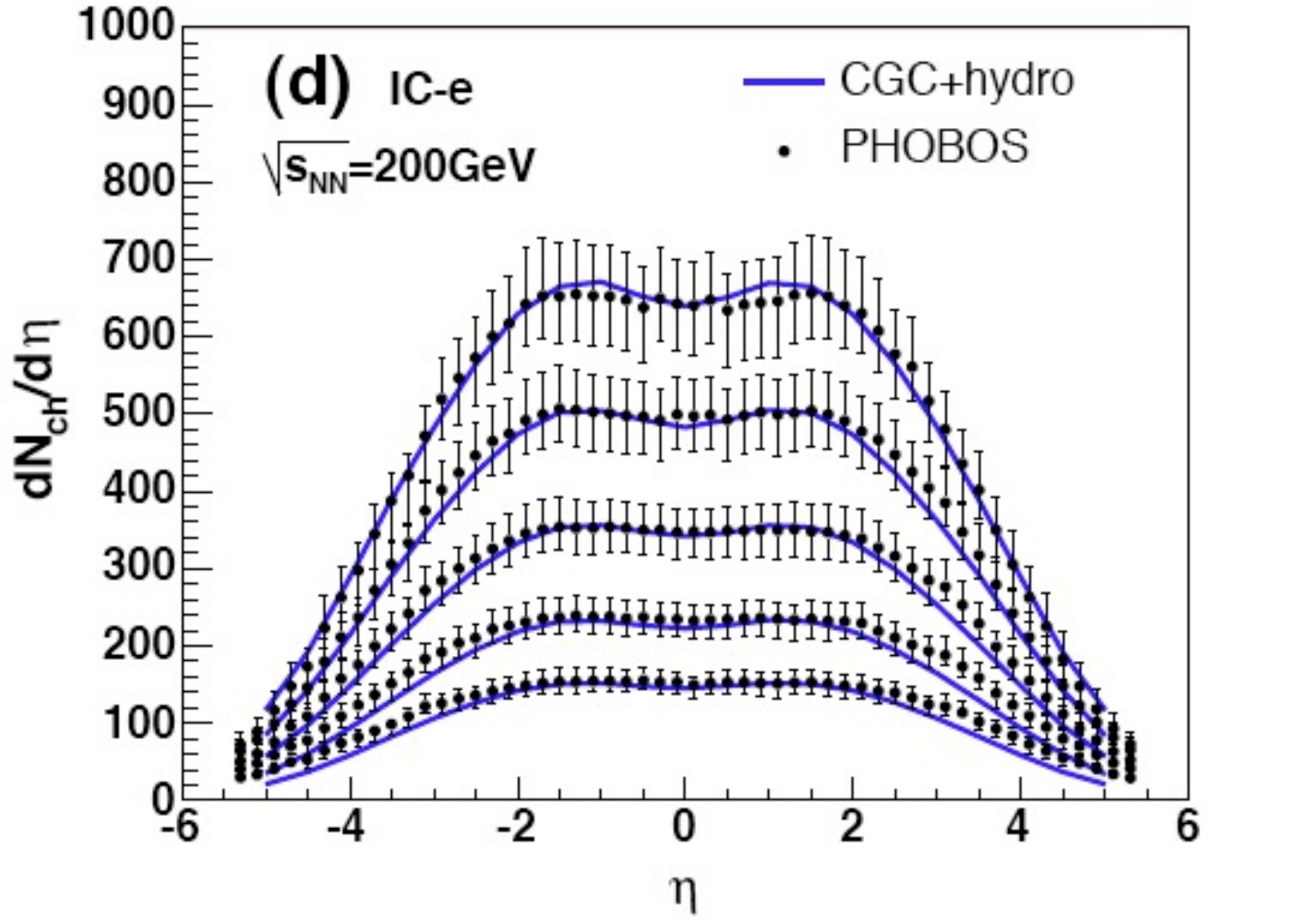} 
\end{center}
\caption{The multiplicity as a function of rapidity in the computations of Hirano and Nara based on CGC-Glasma initial conditions.}
\label{hirano}
\end{figure}

Using such scaling relations for the initial multiplicity of gluons, and a hydrodynamic description of
the evolution of the matter produced in such collisions, the CGC-Glasma ideas can be used to predict the total multplicity as a function of rapidity and centrality.  The results of such a computation\cite{hirano} are shown in Fig. \ref{hirano}

Elliptic flow correlates the anisotropy of the momentum space-distribution of particles with respect to the reaction plane of a non-zero impact parameter collision with the anisotropy of the matter distribution
whose overlap generates the collision.
The CGC-Glasma provide sharper edges for nuclei than that predicted by the standard Glauber type initial conditions.\cite{flow}  This  means that the flow generated would be larger from the Glasma if all other things were equal.  
The issues that are difficult to resolve at this time are the effects of finite viscosity.  If viscosities are very small, then the initial conditions associated with the  CGC would over-predict elliptic flow.  For moderate viscosity however, the CGC-Glasma provides a good description.  Perhaps this might be resolved at the LHC.  The flow values measured in central collisions at RHIC are very close to the upper limit predicted by perfect fluid hydrodynamics and Glauber type initial conditions.  If the flow exceeds this at the LHC, then the additional flow might attributed to the CGC-Glasma.

Perhaps the most direct signal of the initial conditions predicted by the Glasma would be long range rapidity correlations.  The longitudinal fields in the Glasma  are uniform over large longitudinal distance scales. In momentum space, this corresponds to long range in rapidity.  If there is a such a long range correlation,  it must be established very early  in the collision,
\begin{equation}
             \tau_i = \tau_f e^{-\Delta y/2}
\end{equation}
In the Glasma, correlations are set up at time scales corresponding to the time it takes the sheets of Colored Glass to pass through one another.  There can be correlations even between valence quarks
of the different nuclei that span the entire rapidity interval.

\begin{figure}[htbp]
\begin{center}
\includegraphics[width=0.99\textwidth]{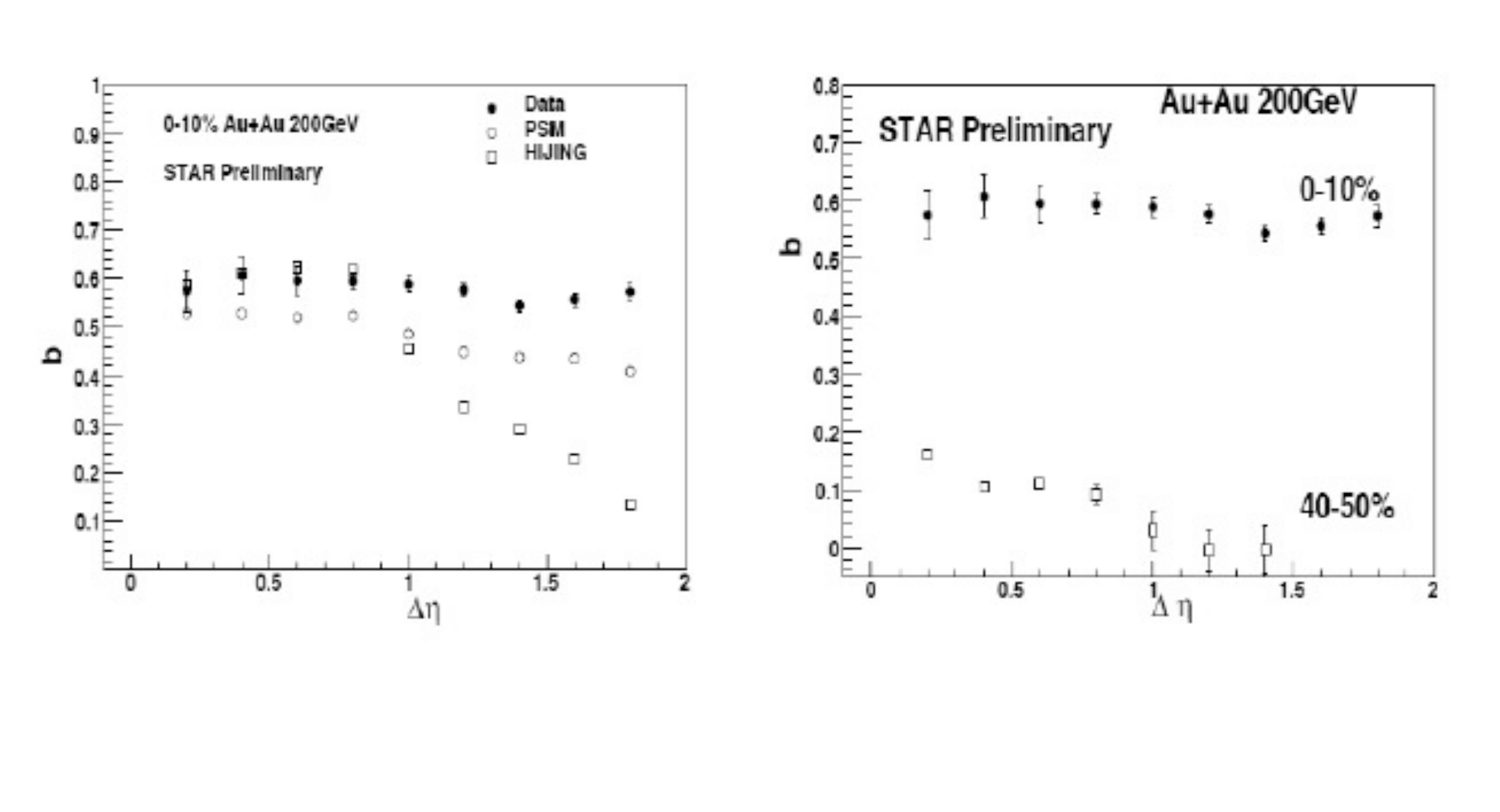} 
\end{center}
\caption{(a) The forward backward correlation for central events as a function of rapidity gap.  The data is compared to HIJING which has no long range correlation and to the Parton String Model, which has
long range correlations similar to those in the Glasma.  (b)  The forward backward correlation as a function of centrality and rapidity gap }
\label{forwardbackward}
\end{figure}

The simplest long range correlation is the forward backward mutliplicity correlation.  This is defined by 
\begin{equation}
  b(y_1,y_2) = {
   {\left< {{dN} \over {dy_1}} {{dN} \over {dy_2}} \right> -  \left< {{dN} \over {dy_1}} \right> \left< {{dN} \over {dy_2}} \right>}
   \over
     {\left< \left({{dN} \over {dy}}\right)^2  \right> -\left< {{dN} \over {dy} }\right>^2 }
    }
\end{equation}
In this equation,  $y$ is typically chosen as the midpoint between $y_1$ and $y_2$.  The gap $\Delta y$
will be chosen as the the absolute value of the difference between the rapidity $y_1$ and the midpoint rapidity. $\mid y_1 - y \mid$  The preliminary results from the STAR experiment presented in Quark Matter 2006,\cite{fb}
are shown in Fig. \ref{forwardbackward}.  There is a huge correlation between the forward-backward
rapidity fluctuations for central events!.  This is far in excess from what is predicted by Monte Carlo event generators such as HIJING which only build in impact parameter correlations.\cite{hijing}  The Parton String Model\cite{psm} does reasonably well, and this model has the long range features of the correlation which are built into
the Glasma description.\cite{glasmafb}

The Glasma provides a qualitative description of the behavior seen in the preliminary STAR data.\cite{glasmafb}.  There is a long range piece generated by the classical fields which is associated with the longitudinal Glasma fields.  The long range piece also gives a nonzero contribution to the fluctuations in rapidity.\cite{dumitru}
This piece is boost independent, and of strength $1/\alpha_S^2$,
because we need two powers of the multiplicity.  There is also a short range correlation piece which
is  higher order in $\alpha_S$.   There is probably a significant contribution to the short range correlation piece from resonance decays, which has no coherence factors associated with it, so could be taken as zero'th order in $\alpha_s$.  In the numerator of $b$ only the long range component contributes at large rapidity separation.  In the denominator, which is at zero rapidity separation, both
the short range and long range piece contribute.  Therefore $b$ is of the form
\begin{equation}
     b = {       {a_{longrange}(\Delta y) }  \over  {a_{longrange}(\Delta y = 0) + a_{shortrange}(\Delta y = 0)} }
\end{equation}
The short range piece is higher order in $\alpha_S$ than is the long range pieces so for more
central collisions, the coefficient $b$ goes to 1. Note that depending upon the relative sign of the long range and short range components, this ratio could approach 1 from above or below as we vary $\Delta y$.\cite{hidaka} To get a precise comparison between the Glasma and experiment, both the long range and short range pieces must be computed, and this has not yet been done.

\begin{figure}[htbp]
\begin{center}
\includegraphics[width=0.99\textwidth]{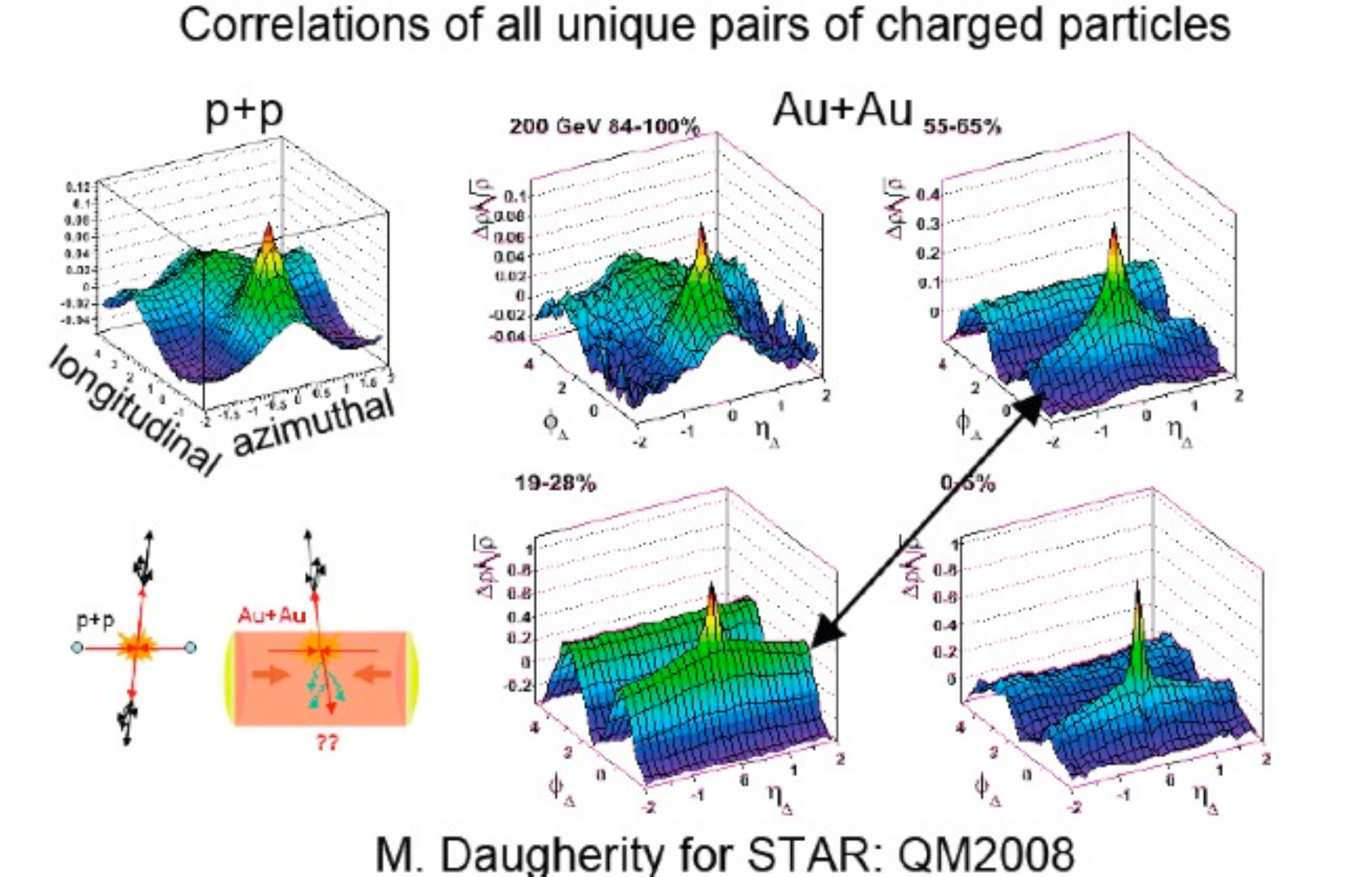} 
\end{center}
\caption{The two particle correlation function as a function of centrailty, azimuthal angle and rapidity,
as presented from preliminary data at Quark Matter 2008. }
\label{ridge}
\end{figure}

Perhaps a more direct observation of the flux tubes of the Glasma comes from the "ridge" phenomena
reported by Star and Phobos.\cite{ridgestar} - \cite{ridgephobos}  This is most dramatically illustrated and most easy to understand for the two particle correlation function where the trigger particle has been integrated over all momentum.  The ridge phenomenon persists at high momentum for the trigger particle as well, but to understand this one needs to understand the effects of jet quenching on the trigger particle.  The two particle correlation is shown in Fig. \ref{ridge}.
In the plot, one clearly sees a short range correlation and a long range correlation in rapidity.  There is peaking in azimuthal angle.  This long range component looks like the image of a flux tube.  
 
 To see how this arises, first note that in the decay of a flux tube
 \begin{equation}
           a(y_1,y_2) =  {{<dN/dy_1dy_2> - <dN/dy_1><dN/dy_2>} \over <\left(dN/dy\right)^2> }
\end{equation}
should be proportional at large rapidity gap to 
\begin{eqnarray}
         a & \sim &{1 \over {Number~of~Flux~Tubes}} \nonumber \\
         & \sim & { r^2_{fluxtube} \over R^2_{hadron}} \nonumber \\
         & \sim & {1 \over {Q_{sat}^2 R^2_{hadron}}}
\end{eqnarray}
The experimental quantity  plotted is
\begin{equation}
         A = a {{dN} \over {dy}} \sim {1 \over \alpha_S(Q_{sat})} 
\end{equation}

To compare with experiment, one needs the angular distribution which is also flat in the restframe of the flux tube.  This distribution is modified by the transverse expansion of the QGP and hadronic gas.
Using parameters determined by fits to $p_T$ spectra using the blast wave model,
one can predict both the dependence of the ridge amplitude on centrality and the angular depence. 
The factor of $\alpha_S$ is determined by the fits of Kharzeev and Nardi to the total multiplicity.\cite{kharzeev} This is essentially a one parameter fit, the overall normalization of the amplitude at one centrality, and fits the STAR data very well.

In addition to the ridge phenomena, there are forward backward correlations which can be determined as a function of centrality.  There should be some suppression of associated jets produced in the backward direction due to the CGC in $dA$ collisions.\cite{marquet}  This may have been seen in the STAR experiment.\cite{bland}

\section{Thermalization of the Glasma}

The decay of the flux tubes in the Glasma may be computed from solving the Yang-Mills classical equations, with initial conditions determined by the Color Glass Condensate. These equations are well defined, and starting from field configuration independent of rapidity, one generates solutions at late time which are also independent of rapidity.  These configurations do not look at all like thermalized configurations.  Nevertheless thermalization might occur by scattering of particles, and there have been
scenarios developed for this.\cite{son}

There are a number of recent speculations concerning the thermalization of the Glasma.  Their common feature is that they require plasma instabilities or turbulence.\cite{mrocz}-\cite{fuji}  To understand how this might come about, we need to first remember that the Glasma fields come from longitudinal fields.  The strength of these fields is independent of their space time rapidity.
The fields are however random on a transverse size scale of order $1/Q_{sat}$
This means that the longitudinal momentum of these fields will be small compared to the transverse part,
since the typical momenta in the fields is ultimately tied to the scale of inhomogeneity.

Non-linear systems which initially are restricted to some regions of phase space, ultimately fill up phase space.  This is a consequence of ergodicity.  This typically happens exponentially rapidly through the development of turbulence.  The simplest example of this phenomena is if one has a time dependent field homogeneous in space.  The classical field theory is free field theory plus very small interaction terms.  If one adds to the system very small spatially inhomogeneous perturbations, these perturbations grow exponentially rapidly, and become large, and their value at any time depends sensitively on the initial conditions.  This is a consequence of parametric resonance amplification, and provides a model for reheating from a coherent axion or inflaton field in cosmology.

In the Glasma, the field starts as a field uniform in rapidity.  If one adds small fluctuations which are
random in rapidity, they in fact exponentially grow.\cite{mrocz}-\cite{fukushima}  The characteristic time scale corresponds to exponential growth in the square root of time with a time scale $t \sim 1/Q_{sat}$.  During this time, the fields are still large and coherent.  This means that even small fluctuation fields interact with them strongly,
since the interactions, of order $g$, multiply strong fields with strength of order $1/g$.  It is again coherence which drives the system.

Ideally one would like the system to become thermalized and get into the hydrodynamic expansion mode described by Bjorken.\cite{shuryakanishetty}-\cite{bjorken}  From that point on, the matter is described as a QGP.  Perhaps the most dangerous time to make this transition is, assuming the turbulence occurs, when the fields are dilute, but the time is before $1/(\alpha_S^2 Q_{sat})$, which is the naive thermalization time.  Perhaps this is resolved since when the fields become dilute, their interactions become strong, and perhaps there is a transition to a strongly interacting quark gluon plasma.

\section{Conclusions}

One of the outstanding problems of theoretical physics has been to understand the high energy limit
for strongly interacting paticles.  This was the goal of the parton model.  In an attempt to understand this limit, one has posited the existence of new forms of matter, such as the Color Glass Condensate and the Glasma.  Understanding such matter is of intrinsic scientific interest, and enriches our knowledge of the possible ways that nature may manifest itself, particularly for highly dense strongly interacting systems.  Such knowledge also allows
us to predict the initial conditions for the Quark Gluon Plasma as it may be formed in heavy ion collisions.

\section*{Acknowledgments}
I thank Edmond Iancu and Raju Venugopalan in particular, and my many colleagues
who have participated in the development of the ideas presented here.  I thank Reinhardt Stock
for encouraging me to write up this manuscript.
My research is supported under DOE Contract No.
DE-AC02-98CH10886.

\end{document}